\documentclass{article}
\pdfoutput=1

\usepackage{arxiv}

\usepackage[utf8]{inputenc} 
\usepackage[T1]{fontenc}    
\usepackage{hyperref}       
\usepackage{url}            
\usepackage{booktabs}       
\usepackage{amsfonts}       
\usepackage{nicefrac}       
\usepackage{microtype}      
\usepackage{lipsum}		
\usepackage{graphicx}
\usepackage[square,numbers]{natbib}
\usepackage{doi}
\usepackage{natbib}
\usepackage{caption}

\title{Vaccine Discourse on Twitter During the COVID-19 Pandemic}

\author{Gabriel ~Lindelöf\\
	Department of Computer Science\\
	Aalto University\\
	02150 Espoo, Finland \\
	\texttt{gabriel.lindelof@aalto.fi} \\
	\And
	Talayeh ~Aledavood \\
	Department of Computer Science\\
	Aalto University\\
	02150 Espoo, Finland \\
	\texttt{talayeh.aledavood@aalto.fi} \\
        \And
    Barbara ~Keller \\
	Department of Computer Science\\
	Aalto University\\
	02150 Espoo, Finland \\
	\texttt{barbara.keller@aalto.fi} \\
}

\hypersetup{
pdftitle={Vaccine Discourse on Twitter During the COVID-19 Pandemic},
pdfsubject={cs.CY},
pdfauthor={Gabriel ~Lindelöf, Talayeh ~Aledavood, Barbara ~Keller},
pdfkeywords={COVID-19 vaccines, SARS-COV-2, vaccine hesitancy, social media, twitter, natural language processing, machine learning, stance detection, topic modeling},
}

\begin{document}

\maketitle

\begin{abstract}
Since the onset of the COVID-19 pandemic, vaccines have been an important topic in public discourse. 
The discussions around vaccines are polarized as some see them as an important measure to end the pandemic, and others are hesitant or find them harmful. A significant portion of these discussions take place openly on social media platforms. This allows us to closely monitor the opinions of different groups and their changes over time. This study investigates posts related to COVID-19 vaccines on Twitter and focuses on those which have a negative stance toward vaccines. We look into the evolution of the percentage of negative tweets over time. We also examine the different topics discussed in these tweets, in order to understand the concerns and discussion points of those holding a negative stance toward the vaccines. A dataset of 16,713,238 English tweets related to COVID-19 vaccines was collected covering the period from March 1, 2020, to July 31, 2021. We used the Scikit-learn Python library to apply a support vector machine (SVM) classifier to identify the tweets with a negative stance toward the COVID-19 vaccines. A total of 5,163 tweets were used to train the classifier, out of which a subset of 2,484 tweets were manually annotated by us and made publicly available. We used the BERTtopic model to extract and investigate the topics discussed within the negative tweets and how they changed over time. We show that the negativity with respect to COVID-19 vaccines has decreased over time along with the vaccine roll-outs. We identify 37 topics of discussion and present their respective importance over time. We show that popular topics consist of conspiratorial discussions such as 5G towers and microchips, but also contain legitimate concerns around vaccination safety and side effects as well as concerns about policies. The most prevalent topic among vaccine-hesitant tweets is related to the use of mRNA and fears about speculated negative effects on our DNA. Hesitancy toward vaccines existed prior to COVID-19. However, given the dimension and circumstances surrounding the COVID-19 pandemic, some new areas of hesitancy and negativity toward the COVID-19 vaccines have arisen, for example, whether there has been enough time for them to be properly tested. There is also an unprecedented amount of conspiracy theories associated with them. Our study shows that even unpopular opinions or conspiracy theories can become widespread when paired with a widely popular discussion topic such as COVID-19 vaccines. Understanding the concerns and the discussed topics and how they change over time is essential for policymakers and public health authorities to provide better and in-time information and policies, to facilitate vaccination of the population in future similar crises.

\end{abstract}


\newpage

\section{Introduction}
Since the emergence of the COVID-19 pandemic, vaccines against the SARS-COVID virus have become a highly salient topic in public discourse. While most agree that the pandemic should end as soon as possible, the opinions on how and through which mechanisms or policies to get there, differ greatly. As a main point of disagreement, many see vaccination (and re-vaccination) of the majority of the world's population as the only way to have the pandemic fully under control, while another group of people is hesitant or completely opposed to the idea of getting vaccinated. The importance and sensitivity of the topic of vaccination leads to a large amount of discourse, commonly expressed on social media platforms, which are often highly polarized. Under circumstances such as social distancing and remote work, imposed under the pandemic, social media platforms have taken an even more central role in people's lives \cite{ijerph18010014}. In these online social spheres, people openly discuss and share their opinions on vaccines with each other globally. A better understanding of how these conversations have developed over time and more insight into the topics discussed within them, can help us better understand those hesitant to the vaccines or those with low confidence in the surrounding processes.

Many attempts have been made to define the concept of vaccine hesitancy \cite{MACDONALD20154161}. One such attempt was made by the SAGE working group on vaccine hesitancy who proposed the following definition \textit{``Vaccine hesitancy refers to delay in acceptance or refusal of vaccines despite availability of vaccine services. Vaccine hesitancy is complex and context-specific, varying across time, place, and vaccines. It is influenced by factors such as complacency, convenience and confidence''} \cite{sage_report_2015}. A word commonly used in relation to hesitancy is confidence, which highlights the aspect of trust in vaccines and trust in the actors involved in the process, such as health care workers, researchers, governments, and pharmaceutical companies \cite{MACDONALD20154161}. The current study focuses on tweets expressing a negative stance toward the COVID-19 vaccines. This includes those tweets expressing a hesitancy toward taking available vaccines, a negative attitude toward policies promoting vaccination, or distrust in actors involved in the vaccination process. We explore how much space these voices take on social media and what the most prevalent topics of discussion are.

A better understanding of individuals with low trust in vaccines or a hesitancy toward taking a vaccine can help shape interventions to improve trust. This will be important to get the current COVID-19 pandemic fully under control, but can also contribute valuable insights how to shape communication in similar situations in the future. Although previous studies have looked at the discourse around COVID-19 vaccines on Twitter \cite{kunneman_monitoring_2020, yousefinaghani_analysis_2021}, the current study is one of the first to cover the entire period from the WHO declaration of the pandemic until the summer of 2021. The study will also provide temporal insights into how these discussions have changed over time. 

\textit{Our main contributions are as follows:}
\begin{enumerate}
  \item A classifier capable of identifying tweets that express a negative stance toward COVID-19 vaccines.
  \item A timeline of the development of the percentage of tweets expressing a negative stance toward COVID-19 vaccines for the first 18 months of the COVID-2019 pandemic.
  \item An overview of the major topics discussed by those with a negative stance during this time, their development, and what events co-occurred with changes in the discourse. 
  \item A dataset of 2,484 tweets manually labeled with their stance toward the COVID-19 vaccines, along with the codebook used in the annotation process \cite{dataset}.
\end{enumerate}

\section{Background}
\subsection{Sentiment Analysis }
A number of approaches have been taken to analyze sentiments around infectious diseases on social media \cite{alamoodi_sentiment_2021}. Sentiment mining on social media has proven to be a valuable resource to understand people's opinions about ongoing events, and could potentially help with controlling pandemics \cite{singh_sentiment_2018}. Previous studies investigating sentiments around infectious diseases can broadly be grouped into three categories; lexicon-based, machine learning, and hybrids between the former two \cite{alamoodi_sentiment_2021}. Before the dawn of the COVID-19 pandemic \citet{du_optimization_2017} used a machine learning approach to investigate the stance toward HPV vaccines on Twitter. Using an annotated dataset of 6,000 tweets they were able to train an SVM classifier that could classify tweets as {positive}, {negative}, or {neutral} with satisfactory performance.

At the beginning of the COVID-19 pandemic \citet{medford_infodemic_2020} investigated sentiments discussed on Twitter. They found that approximately 50\% of COVID-19 related tweets could be classified as showing fear, while 30\% were expressing surprise. Dominating topics discussed were economical and political impacts of the virus, quarantine efforts, as well as transmission of the virus and how to prevent it. 

\subsection{Stance Detection}
Stance detection analyzes textual input from people and identifies whether someone is favorable, against, or neutral toward a certain topic. Stance detection is related to sentiment analysis but is known to be a more difficult task \cite{skeppstedt-2017-automatic}. Sentiment analysis aims to identify the opinion of a person and determine whether the text bears positive, negative, or neutral sentiments. A negative sentiment of a text does not always imply an unfavorable stance toward the topic. As an example, \citet{skeppstedt-2017-automatic} uses the phrase ``The diseases that vaccination can protect you from are horrible.''. This phrase is favorable toward the pre-chosen topic of vaccination but contains negative sentiments toward an identified topic which is ``diseases''. However, in previous literature, there is not always a clear distinction between stance detection and sentiment analysis. For instance, some of the work that study the stance of social media users toward vaccination does not use the term stance \cite{glandt2021stance}.

Stance detection has also been used by \citet{cotfas_covid-19_2021} to study hesitancy during the month following the UK vaccination roll-out. Their cleaned dataset contained ~1.2 million tweets and approximately 7 out of 10 were classified as neutral, and 2 out of 10 as having a negative stance. Some of the larger peaks in volume of negative tweets correlated with the roll-out of the Pfizer vaccine (Dec 8, 2020), FDA approval of Moderna (Dec 17, 2020) and the vaccination dry-run in India (Jan 2, 2021) \cite{cotfas_covid-19_2021}. Some of the most discussed topics were labeled as being about mistrust, scam, side effects, and the hiding of relevant information. The period before or after the month of the roll-out was not studied, nor was it investigated how the popularity of found topics developed over time.

\subsection{Vaccination Discourse on Social Media}
The widespread use of social media gives us the opportunity to gain insights about opinions of a broad population. A topic of particular interest with respect to public health is the discourse around vaccination. Understanding concerns and insecurities about vaccinations can help policymakers in finding an appropriate way to address them. In 2011 \citet{Sal2011} assessed the sentiments expressed in tweets toward vaccination in the fall wave of the H1N1 (swine flu) pandemic. The authors could establish a correlation between the expressed sentiments on Twitter and the corresponding estimated vaccination rate (gained via phone surveys) in the same area. On the other hand, another study on vaccination discourse on Facebook, focusing on the common flu vaccination, found an asymmetric participation of vaccination defenders and critics \cite{Garg2020}. While the defenders were able to reach 24\% of the investigated network, the vaccination critics were only able to reach 8\%.

With the onset of the COVID-19 pandemic, extensive work was done on the area of COVID-19 vaccinations. Some studies focus on the demographics of the vaccination hesitant population \cite{Malik2020,Thunstrom2020} while others try to understand what leads to this hesitancy \cite{Bonnevie2021, poddar2021}, finding similar to our study, that common concerns are potential side effects and distrust toward the pharmaceutical industry. \citet{Melt2021} used a similar approach as us but focused on the social media platform Reddit. They investigated topics of vaccine discussion for the period of December 1, 2020, to May 15, 2021 and found that a majority of posts were using a positive tone. \citet{lyu_covid-19_2021} focused on tweets made about COVID-19 from the declaration of the pandemic until February 2021. Their results showed that opinions and emotions around vaccines was the most tweeted topic, and that sentiment became more positive over time.

\section{Materials and Methods}

\subsection{Dataset}
\label{section:dataset}

Tweets were collected using the academic research track of the Twitter Application Programming Interface (API).
Using the full-archive search we were able to retroactively collect Tweets from each day for the period of March 1, 2020, to July 31, 2021. We queried English-speaking tweets containing synonyms of the words ``COVID-19'' and ``vaccine'' excluding any retweets. A Python script utilizing Twarc2 \cite{twarc2} was used to send requests and collect results.

Since random sampling is not possible when gathering historical tweets, tweets were collected at six different time points of each day, corresponding to noon in six major time zones with a large population of English speakers: AEST, IST, CET, EST and PST. This means we get a more balanced dataset collected during high activity hours for different parts of the world. 

\begin{figure*}[h]
  \centering
  \includegraphics[width=\linewidth]{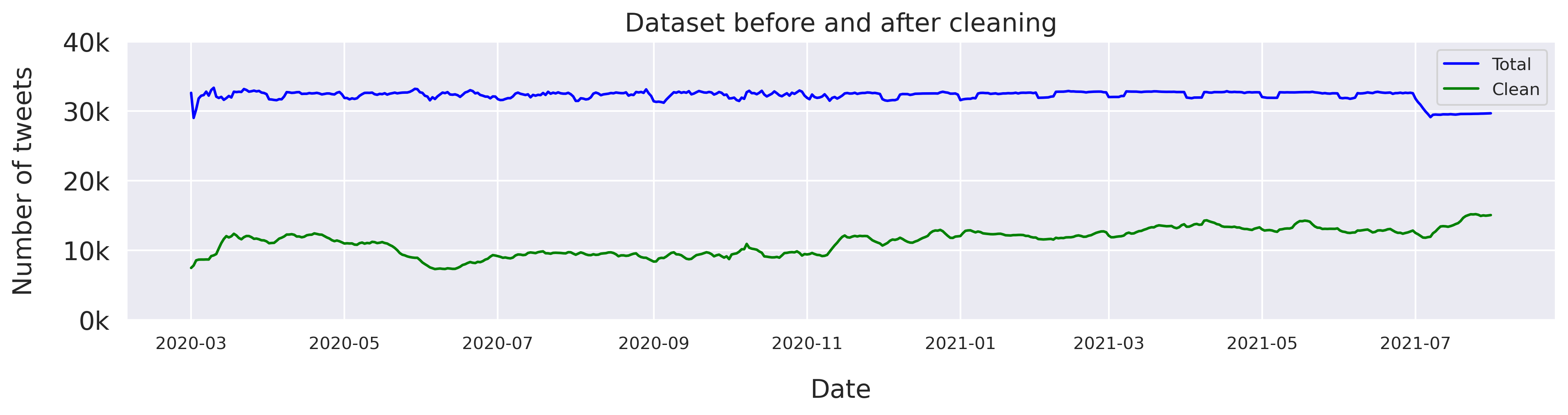}
  \caption{7-day moving average of tweets before and after the cleaning steps described in \ref{section:dataset}.}
  \label{figure:cleaning}
\end{figure*}

At each time point, 30 twitter-pages of tweets were collected. Since pages do not necessarily correspond to a fixed number of tweets, there are some variations in the collected volume from one day to another, as can be observed in Figure \ref{figure:cleaning}. For the first day of each month, as well as all days in the month of July 2021, the API unexpectedly returned a lower number of tweets per page. Given the large number of collected tweets for each day we did not consider this an issue for the analysis. 

The collected dataset contained a total of 16,713,238 tweets, with an average of 32,203 for each day in the period. To put these numbers into perspective we also investigated how many tweets matched our search query in total. Using the \textit{Counts endpoint} of the Twitter API we can quickly retrieve how many tweets match our keywords, without having to go through the lengthy process of collecting them all. During the investigated period a total of 85 million were made matching the query, meaning our sample represents roughly 20\% of relevant tweets.

As we were interested in the discourse led by people, tweets posted by the bots active on Twitter needed to be excluded. To investigate their prevalence in the dataset a random sample of 300 tweets were run through the OSoMe Botometer API \cite{Sayyadiharikandeh_2020}. Botometer returns a \textit{complete automation probability} (CAP) that indicates the likelihood of an account being automated. 26\% of tweets were made from accounts with a CAP above .80, and out of these accounts 87\% were sharing links. Shared links pose their own challenges in classification tasks. On one hand, it is difficult to judge if the link shared does indeed reflect the stance of the sharer and on the other hand retrieving the information following the link can be difficult especially as the information can change over time. Based on these challenges and the fact that many bots are sharing links, we decided to remove all tweets containing links. After this, 7,292,705 tweets remained, corresponding to 44\% of the dataset. Tweets with identical text content to other tweets from the same author were also removed in order to reduce the amount of spam, making the new total 5,966,905. Finally, to reduce over-representation of individual users, tweets made from users with more than 1000 tweets contained in this dataset were removed. An example of such cases were bots that continuously announced free vaccination appointments. This removed 108,749 tweets, making the total number of tweets in the cleaned dataset 5,858,156. 

\subsection{Pre-Processing} 
The text data of the collected tweets were extracted and transformed using a few common pre-processing steps. Text was lower-cased, special characters were removed, ``{\&}'' symbols were replaced with ``and'' and letters occurring more than twice in a row were removed (helllo becomes hello). Text was also separated into units of words using the NLTK TweetTokenizer. Stop-words contributing little meaning to the text such as ``the'', ``a'', and ``in'' were removed using the NLTK English stop word list. Since the terms used in the twitter search query occur at least once in every tweet, they add bias and contribute only little additional information, and therefore were also removed. The same pre-processing steps were taken on the text data for both the classifier and the topic model. Pre-processing is not generally recommended for BERTopic (the topic model used in our analysis, see section \ref{sec:tp_negative_tweets}), unless the data contains a lot of noise such as HTML tags. In this case, however, pre-processing led to a better topic model. Perhaps the large volume of data analyzed in this study made the small loss of information due to the cleaning process negligible.

\subsection{Annotation}
A first step to computationally categorizing documents (in our case tweets) into different groups (e.g. negative and not negative stance toward COVID-19 vaccines) is to manually annotate a subsample of the documents with the correct labels as determined by the annotators. A codebook is often used to have a common framework between different annotators. The codebook sets out the rules that help decide which label a tweet receives. Although the classification in this study is limited to a binary negative/not negative, we used three classes for our codebook. This made the annotation more intuitive, and also enables future studies to make a distinction between the neural and positive stances when required. This dataset which was annotated using the codebook in Table \ref{table:codebook}, is published along with this article, so that other researchers can benefit from it. In our codebook a tweet can have a positive, negative or neutral/unclear stance toward the COVID-19 vaccines. The explanations of these categories were inspired by a previous codebook classifying tweets about HPV vaccines \cite{du_optimization_2017}. 

\begin{table}[]
    \centering
    \captionsetup{justification=centering}
  \caption{The codebook developed for the purpose of this study with example tweets for each category.}
\begin{tabular}{p{1.1cm}|p{9cm}|p{4cm}}
Stance & Definition & Example tweet \\ \hline
\vspace{5pt} Positive &\vspace{2pt} Showing positive opinion toward the COVID-19 vaccines. 

Prompting uptake of the vaccines. 

Expressing the intention of taking or having taken the vaccine. &  \vspace{2pt}``Im getting vaccinated tomorrow yay'' \\  \hline
\vspace{15pt} Negative & \vspace{2pt}Expressing concerns around safety, efficacy, injuries, cost, resistance to COVID-19 vaccines due to cultural or emotional matters. 

Discouraging uptake of the vaccines. Expressing intention of not taking or refusing the vaccines. 

Questioning motives behind vaccine deployment, for example, those of scientists, pharmaceuticals companies or governments. & \vspace{2pt}``No mask and no COVID19 vaccine for me!'' \\ \hline
\vspace{35pt} Neutral & \vspace{2pt}Contain no stance, or stance is unclear. 

Expresses both pro and anti-stance at the same time. 

Expresses stance of someone else without own added input. Not related to COVID-19 vaccine topic. 

Discussion on other medical treatments without relating it to vaccines.

Discussion on vaccines for other diseases. Unclear what the person is trying to say. & \vspace{2pt} ``Would you take a COVID vaccine?''
\end{tabular}

  \label{table:codebook}
\end{table}

Tweets to be annotated were randomly sampled from the cleaned dataset. The first 485 tweets were coded by the authors while the codebook was still being refined. Thereafter two university students were hired to annotate 1999 more tweets using the developed codebook. Out of these, a total of 543 tweets were labeled positive, 471 negative, and the remaining 1,450 as neutral/unclear. All inter-annotator disagreements were discussed in an attempt to find agreement on a final label. If no agreement could be made, the label was set to neutral/unclear. The majority of disagreements were between neutral/unclear and the two other labels. There were only a total of 22 disagreements were one annotator said positive and the other negative.

\subsection{Classifier}
\label{section:classifier}
To investigate the discourse of those with a negative stance toward the vaccines, we had to find a way to identify tweets belonging to this category. Considering the size of the dataset we opted for a machine learning approach. While there are many viable classification algorithms available, previous research classifying stance toward vaccines on Twitter have shown success using support vector machines (SVM) \cite{du_optimization_2017, kunneman_monitoring_2020, cotfas_covid-19_2021}. Initial exploration comparing multinomial naive bayes, random tree and an SVM classifier also showed most promising results for the SVM, leading us to choose this as our classification method. 

SVM is a type of classification algorithm that can be used to automatically divide documents into categories \cite{Noble2006}. In this study, each tweet is a document that needs to be labeled with an either \emph{negative} or \emph{not negative} stance. Using thousands of tweets annotated by hand, the SVM can learn from these examples (training dataset) to automatically label the remaining millions of tweets. The pre-processed tweets were vectorized using \textit{term frequency-inverse document frequency} (TF-IDF), giving each word a weight indicating how important it is in the text. The tweets that had been manually annotated were used to train the SVM to recognize tweets with a negative stance. The training data consisted of 2,484 tweets annotated for the purpose in this study as well as 2,679 annotated tweets made available by \citet{cotfas_covid-19_2021}. Randomized search with cross-validation was used to find optimized parameters for the vectorizer and the SVM. Since the focus was on negative tweets, it was a priority to minimize the number of tweets incorrectly labeled to this category. By prioritizing the precision on negative tweets we can decrease the risk of themes extracted not corresponding to negative tweets \cite{kunneman_monitoring_2020}. We therefore opted for a classifier with the highest precision on the negative class.

\subsection{Topic Modeling} \label{sec:tp_negative_tweets}
Topic modeling is a technique that is used to extract themes from a set of text documents. The current study aims to investigate topics of discourse with a negative stance toward the COVID-19 vaccines. Manually categorizing all 296,321 negative tweets into topics would have been incredibly time-consuming, but topic modeling allows this to be done automatically in a matter of hours. We used BERTopic to train a model and label each tweet with a topic. BERTopic is a topic modeling technique that utilizes a complex language model (BERT) to cluster documents based on semantic similarity. For this work we used the Python implementation of this model available as a package \cite{grootendorst2020bertopic}. A BERT model pre-trained on a large set of texts can be fine-tuned to be used in a wide variety of language recognition tasks \cite{devlin2019bert}. We used the BERTopic default embeddings model that has been fine-tuned on 1B English sentence pairs. 

The model was fitted on all tweets classified by the SVM as negative. BERTopic allows the user to specify the desired number of topics, as well as the minimum number of documents that should constitute a topic. For the purpose of keeping the size of the model manageable, the minimum topic size was set to 500 documents, and the number of topics to ``auto''. Words with a term frequency lower than 0.0001 were also removed. The resulting model contained 37 distinct topics, presented in the results section. 

To explore changes in negative discourse over time, a dynamic topic model was also developed using BERTopic's topics over time function. BERTopic has a parameter that controls the number of topic representations each topic should have in the timeline. A higher value gives more timestamps in the graph, but risks decreasing the quality of the topic representations. This value was set to 35 as this number led to graphs with an appropriate level of granularity for our analysis. To allow for easier comparisons between topics of different sizes the frequencies were normalized to values between 0 and 1. A topic having a higher peak than another in the graphs should therefore not be interpreted as it being more popular, but rather as relative popularity for that particular topic. 

\section{Results}
\label{section:reslts}
In the following section we present the performance of the classifier and the findings based on categorization of all the tweets in the clean dataset using the classifier. We show how negativity toward the COVID-19 vaccines developed over the course of the pandemic, which topics made up this negative discourse, and how these individual topics evolved over time. 

\subsection{Classifier Performance}
The best classifier found using randomized search is presented in Table \ref{table:svm}. A macro F1-Score of 0.67 was achieved with a precision on the negative class of 0.8. This classifier used a RBF kernel and a vectorizer with 3000 features of unigrams and bigrams. The classifier was used to label the 5,858,156 tweets in the cleaned dataset as having a negative or not negative stance toward the COVID-19 vaccines. A total of 296,321 tweets were classified as negative, corresponding to 5.1\% of all clean tweets.

\begin{table}[h]
  \centering
  \captionsetup{justification=centering}
  \caption{Performance of SVM classifier.}
  \label{table:svm}
\begin{tabular}{lllll}
 & Precision & Recall & F1-Score & Support \\
 \hline
Negative      & 0.80 & 0.33 & 0.46 & 131 \\ \hline
Other         & 0.81 & 0.97 & 0.88 & 386 \\ \hline
Accuracy     & & & 0.81 & 517   \\ \hline
Macro avg.    & 0.80 & 0.65 & 0.67 & 517 \\ \hline
Weighted avg. & 0.81 & 0.81 & 0.78 & 517 \\ \hline
\end{tabular}
\end{table}

\subsection{Percentage of Negative Tweets Over Time}
\begin{figure*}[h]
  \centering
  \includegraphics[width=\linewidth]{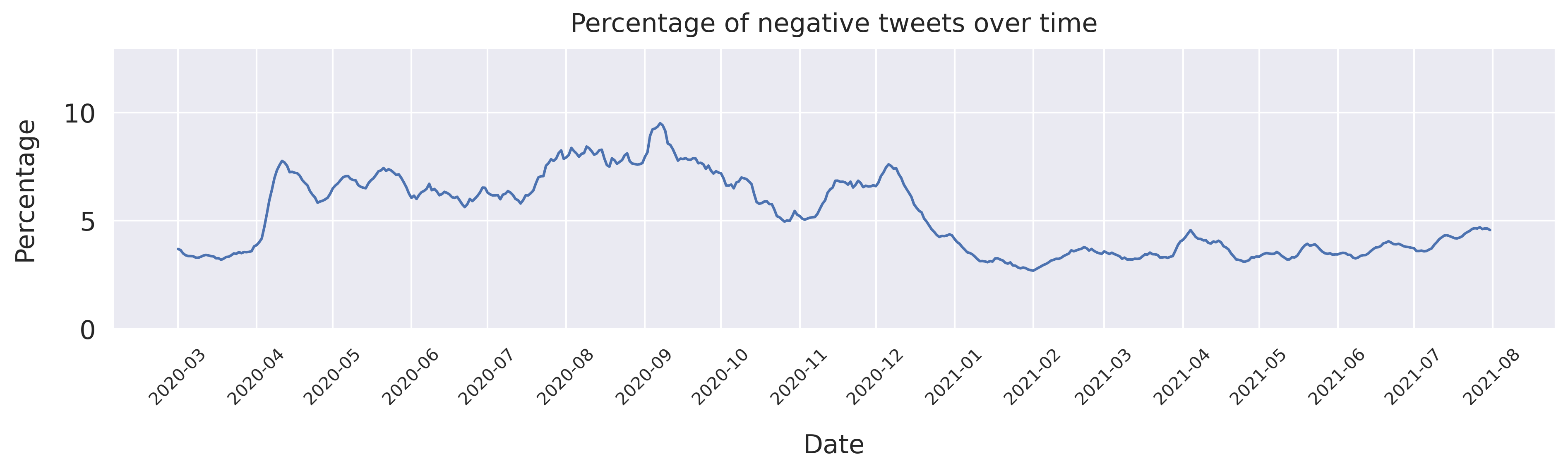}
  \caption{7-day moving average of percentage tweets classified as negative.}
  \label{figure:negative}
\end{figure*}

A timeline of the percentage of tweets with a negative stance toward the COVID-19 vaccines during the period March 1, 2020 to July 31, 2021 can be seen in Figure \ref{figure:negative}. The average percentage of negativity is 5.1\%. This number is slightly lower than what \citet{cotfas_covid-19_2021} found for the period 8 December, 2020 – 7 January, 2021, where the percentage of tweets against the COVID-19 vaccines was 6.78\%. For this same period, our estimate is 4.7\%. The discrepancy could likely be explained by the design choice of our study to employ a conservative classifier that heavily prioritized not having false positives for the negative class, with the cost of more false negatives. Figure \ref{figure:negative} shows the percentage of negative tweets per day for the entire time period. Around the WHO declaration of the pandemic (March 11, 2020), the amount of negativity was quite stable at 4\%, to then rise in April 2020 and remain relatively high for the rest of the year. Co-occurring with the vaccination roll-out in December is a decline in the share of negative tweets, and the following period showed similar negativity to that observed before the declaration of the pandemic.

\subsection{Topic model of Negative Tweets}
The produced topic model contains 37 different topics with sizes ranging from 531 to 119,196 tweets. The topics are presented in Table \ref{table:topics} along with the top keywords that were calculated to be the most representative of each topic using BERTopic's modified version of TF-IDF. A Dynamic Topic Model was also created representing the popularity of each topic at 35 different points in time of the investigated period. In addition to the graphs included in the results section, the timelines of all topics are available in interactive form on the companion website \cite{companion_website}. 

\begin{table}[htbp]

  \captionsetup{justification=centering}
  \caption{Topics of the model ordered by size: The \textit{Frequency} column contains the number of documents predicted to belong to each topic. \textit{Top-words} contains the top-scoring words for each topic. The first topic \textit{UA} contains unassigned documents that did not fit into any of the other topics. Topic 27 contained only usernames as top-words, now censored.}
\label{table:topics}
  \centering
  \begin{tabular}{ccl}
    \toprule
    Topic&Frequency&Top-words\\
    \midrule
\textbf{UA*} & 103953 & gates, take, dont, bill                      \\ \hline
\textbf{0}   & 119196 & virus, take, flu, dont                       \\ \hline
\textbf{1}   & 10415  & dna, mrna, rna, gene                         \\ \hline
\textbf{2}   & 9431   & pharma, trump, big, trust          \\ \hline
\textbf{3}   & 9378   & mask, masks, wear, wearing                   \\ \hline
\textbf{4}   & 3232   & passport, passports, id, travel              \\ \hline
\textbf{5}   & 3058   & africa, africans, black, african             \\ \hline
\textbf{6}   & 3037   & children, kids, school, risk                 \\ \hline
\textbf{7}   & 2889   & china, chinese, trust, virus                 \\ \hline
\textbf{8}   & 2820   & guinea, pigs, pig, first                     \\ \hline
\textbf{9}   & 2159   & exam, india, indian, oxygen                  \\ \hline
\textbf{10}  & 2048   & pfizer, blood, clots, astrazeneca            \\ \hline
\textbf{11}  & 1931   & virus, coronaviruses, years, take            \\ \hline
\textbf{12}  & 1787   & 5g, bill, gates, us                          \\ \hline
\textbf{13}  & 1749   & bill, microchip, chip, tracking                \\ \hline
\textbf{14}  & 1549   & jab, jabs, experimental, jabbed              \\ \hline
\textbf{15}  & 1363   & hydroxychloroquine, zinc, gates, chloroquine \\ \hline
\textbf{16}  & 1289   & test, testing, untested, rushed              \\ \hline
\textbf{17}  & 1184   & polio, measles, smallpox, pox                \\ \hline
\textbf{18}  & 1089   & russian, russia, putin, trust                \\ \hline
\textbf{19}  & 1028   & hiv, aids, years, 40                         \\ \hline
\textbf{20}  & 969    & sars, sarscov2, years, mers                  \\ \hline
\textbf{21}  & 907    & rushed, first, take, im                      \\ \hline
\textbf{22}  & 874    & poison, poisonous, body, take                \\ \hline
\textbf{23}  & 872    & trust, dont, im, wouldnt                     \\ \hline
\textbf{24}  & 827    & experimental, take, taking, experiment       \\ \hline
\textbf{25}  & 794    & antibodies, antibody, test, natural          \\ \hline
\textbf{26}  & 772    & boris, brexit, eu, borisjohnson              \\ \hline
\textbf{27}  & 737    & username, username, username, username*     \\ \hline
\textbf{28}  & 729    & liability, manufacturers, sue, liable        \\ \hline
\textbf{29}  & 676    & cure, treatment, want, cures                 \\ \hline
\textbf{30}  & 629    & science, scientists, trust, dont             \\ \hline
\textbf{31}  & 624    & recovery, rate, 99, need                     \\ \hline
\textbf{32}  & 614    & lockdown, lockdowns, want, dont              \\ \hline
\textbf{33}  & 611    & hcq, pharma, big, gates                      \\ \hline
\textbf{34}  & 570    & injected, injection, inject, body            \\ \hline
\textbf{35}  & 531    & travel, fly, flying, airlines                \\ \hline
  \bottomrule
\end{tabular}
\end{table}

\begin{figure}[tbp]
  \centering
  \includegraphics[width=20cm]{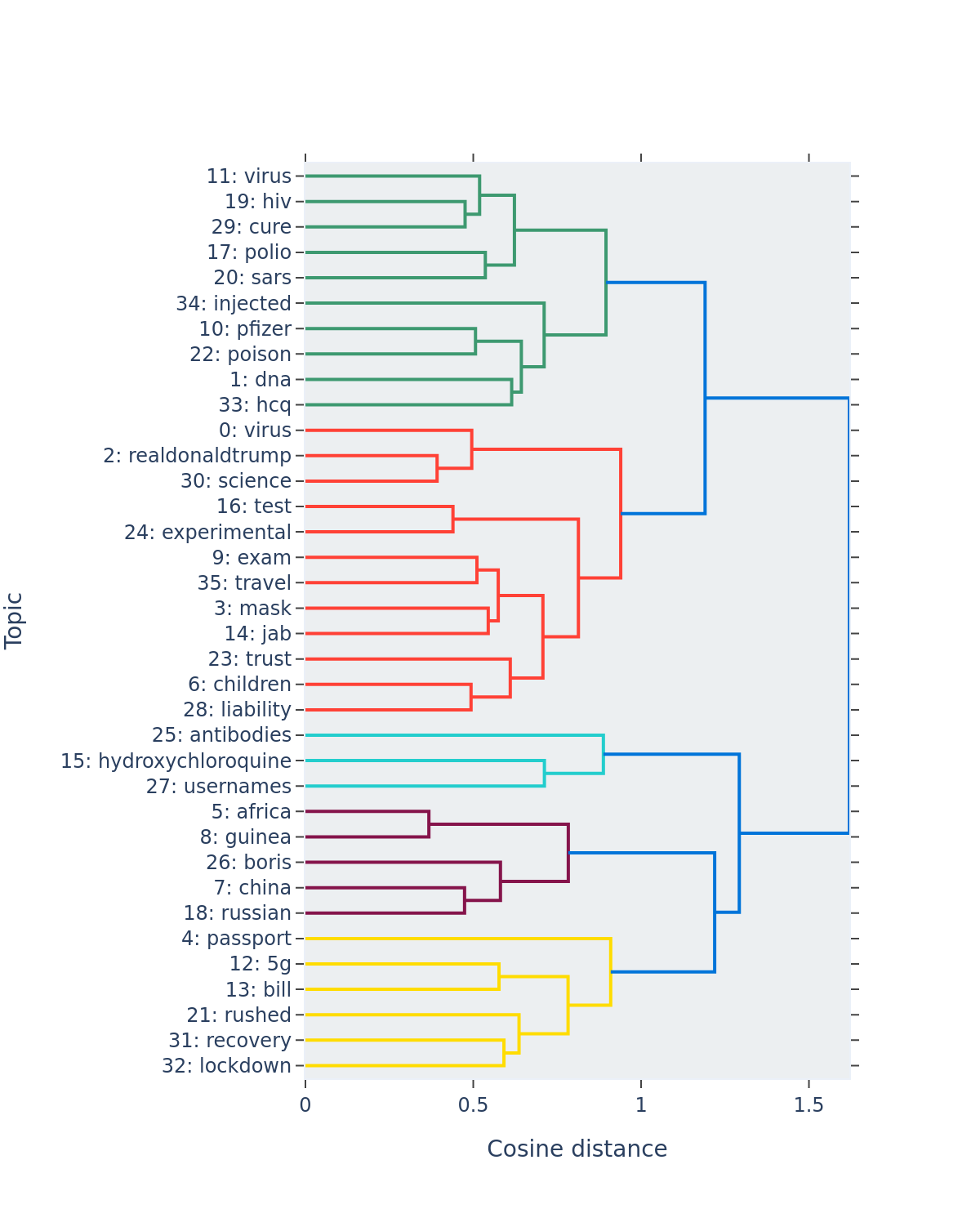}
  \caption{Hierarchical structure of the topics labeled with topic index and the highest scoring word.}
  \label{figure:dendrogram}
\end{figure}

Under this section, a selection of topics from the model will be discussed using example tweets and graphs of their development over time. The topics will be referred to with an index that can be used to locate them in Table \ref{table:topics}. The dendrogram in Figure \ref{figure:dendrogram} shows how closely related topics are, according to our topic model. In the following discussion we have chosen to group some topics together when discussing them, these groupings are based on their closeness in the dendrogram as well as qualitative similarities seen by the authors.

\subsection*{Travel, COVID-passports and territories (topics 4, 7, 9, 18, and 35)}
This group contains topics related to traveling and COVID-passports, as well as discourse around the countries China, Russia, India, and, the continent of Africa.

Topic 4 is a cluster of documents discussing COVID-passports and certificates of different forms as a requirement for travel and other activities. 
The words ``need'' and ``want'' are highly scored in the representation, with many tweets arguing that they don't want or need a vaccine nor a passport. The word mandatory is also prevalent as a result of many discussing the enforcement of such passports. 

\begin{figure*}[h]
  \centering
  \includegraphics[width=\linewidth]{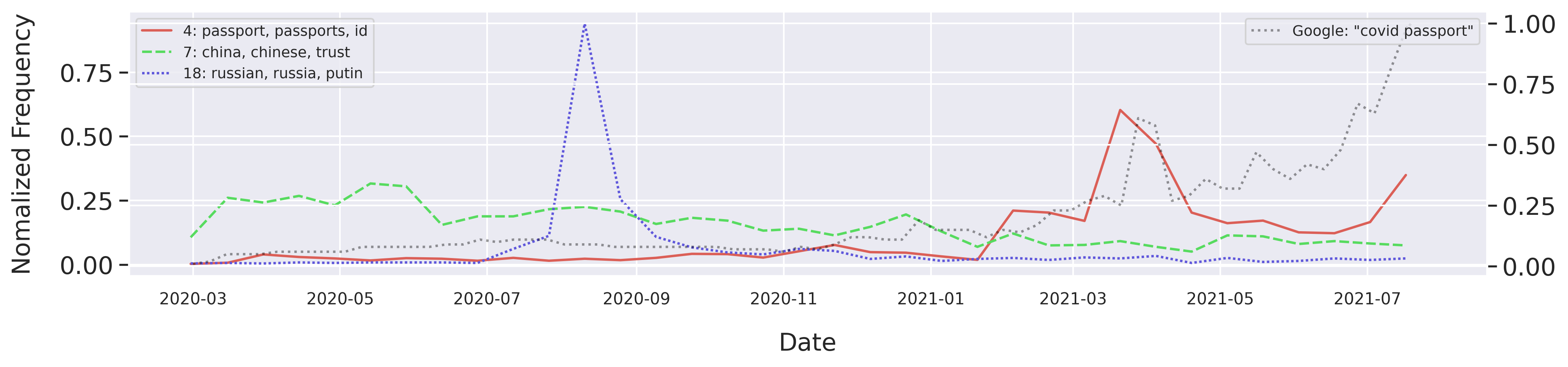}
  \caption{COVID-passports, China, and Russia topic over time. Google Trends normalized search frequency for ``COVID passport'' on the axis to the right. }
  \label{figure:passport_google_china_russia}
\end{figure*}

Figure \ref{figure:passport_google_china_russia} shows that this topic first gained popularity in the negative tweets by the beginning of 2021 as vaccines were being rolled out. This trend is true also for the general search interest as indicated by the blue line showing the Google Trends popularity for the query ``COVID passport''. The topic shows major peaks in both the topic model as well as Google trends, May and July of 2021. The first peak in April has especially high search volume from Great Britain, as indicated when isolating this period in Google Trends. The go-ahead from Boris Johnson for COVID-passports could likely be the event sparking this discussion. Negative tweets from this period discuss COVID-passports in general, but many also speak to Boris Johnson directly. ``@BorisJohnson You must ensure GB does not push vaccine passports. It is intolerable to think Pharma and Gates Foundation should have any influence here. You are elected and accountable-they are not. The vaccinated can still pass on COVID. Stop penalising hard pressed medics to comply. Disgrace.''. Topic 35 about travel contains similar discussion but with a focus on travel, airlines and going abroad, and seems to have had a more constant popularity of discussion throughout the entire period. 

The peak in Figure \ref{figure:passport_google_china_russia} for July is likely related to the European Union's introduction of the EU Digital COVID Certificate (EUDCC). This was introduced to be used in travel within the Schengen area to prove vaccination, recovery or recent negative test. Many tweets in this topic argue the passports are unnecessary and unfair since vaccinated people still can transmit the virus. Others discuss the forging of COVID passports, ``The black market should have a fake vaccine passport or whatever they are calling it.. This will be my approach'', ``Anybody selling a fake vaccine passport yet? Or am I abit too early?''. More conspiratorial voices also discussed how the vaccine passports were a part of a bigger plan: ``No thanks. Vaccine passports where in plan before COVID-19''.

China was also a major topic in the model with 2889 negative tweets being classified to belong to this cluster. Major keywords for this topic is ``chinese'', ``trust'', and ``fake''. Some tweets in this topic express distrust in China, even claiming the virus was ``manufactured'' there:
``I pondered why China has yet to come up with a vaccine for COVID since it began there. Then I remembered that any vaccine would be manufactured in China so why would they hurry? Why would they want to alert the WHO that this virus had to be taken seriously? China is a threat?''. The topic saw an increase in popularity in mid-march 2020 (see Figure \ref{figure:passport_google_china_russia}), when the Chinese vaccine was approved for human testing. Some negative tweets from this period expressed unwillingness to take a Chinese vaccine, while others discussed conspiracies of a China-developed virus. This topic was highly salient in the negative discussion at the beginning of the pandemic and then slowly lost traction over time. 

Russia was also a topic of negative discussion with a majority of tweets between July and November 2020, with a major peak in August, see Figure \ref{figure:passport_google_china_russia}. This peak is likely due to the Russian vaccine Sputnik V being registered August 11. Some of these tweets are discussing whether one should trust a Russian vaccine: ``I wouldn’t want a covid vaccine from putin would you?'' It should be noted that our study only investigates tweets in English and the topics about Russia and China are therefore mainly talking about these countries from a foreign perspective. 

India is also a country prominent in the negative discussion, with a focus on their exams. Most of the discussion in this topic concerns mandatory in-person exams during the pandemic. It should be noted that tweets in this topic may have posed a difficult challenge for the machine learning classifier. That is, some tweets that express negativity toward in-person exams have incorrectly been classified as having a negative stance toward the vaccines. For example, many tweets demanded a shift to online exams, until a vaccine has been made available: ``Shift exams to online mode or no continuous exams. Everyone demanding for Delay in Compliance due dates. Think Same way, students are not punching bag for everyone. We are also human, we are not corona immune. We have not taken corona vaccine yet???''. 
 
The topic about Africa contains several different themes of discussion. A major theme is claims of racist motives behind the vaccines. Users argue that vaccines are being sent to African countries part of a secret plot by Bill Gates to test the side-effects of the vaccines or for the purpose of population control. Both ```bill'', ``guinea'' and ``pigs'' are highly scoring words in the topic representation. Moreover the topic also contain similar claims of racism against African-Americans. This discourse seems to have been sparked by a statement by Melinda Gates where she proposed that the black population should receive priority for the vaccines since they face disproportionate effects from the virus. Another theme of this topic is indicated by the keyword ``madagascar'' in the representation, and concerns a tea that is being used by the Madagascan population with claimed benefits against COVID-19, which some argue makes a vaccine redundant.

\subsection*{Pharma and Alternatives to Vaccines (topics 2, 15, 25, 28, 31, and 33)}
\begin{figure*}[h]
  \centering
  \includegraphics[width=\linewidth]{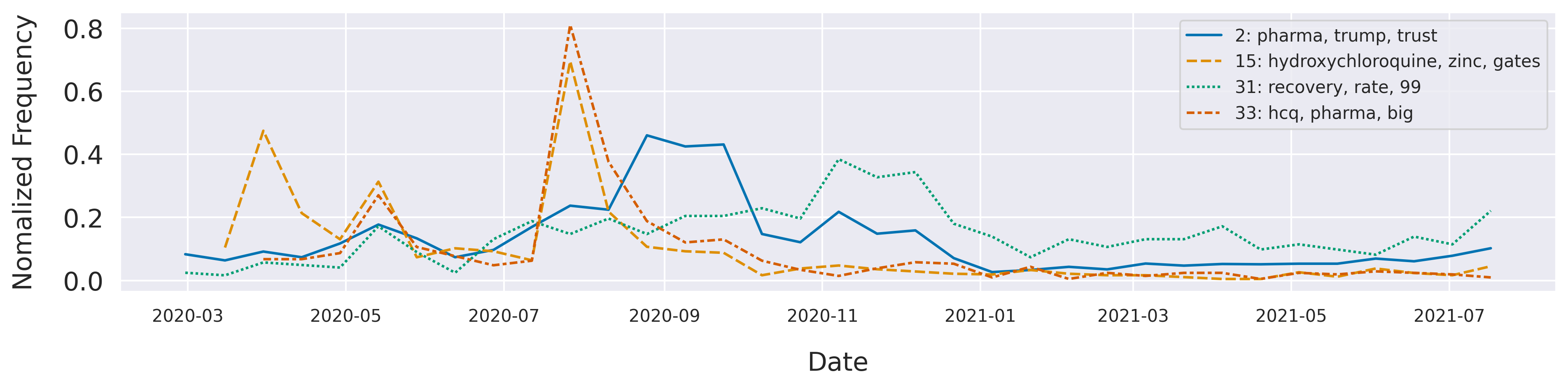}
  \caption{Hydroxychloroquine, pharma, and recovery topics over time.}
  \label{figure:pharma_hcq_99_hcq}
\end{figure*}

Under this heading, we have grouped topics that concern discussion around pharmaceutical companies, their profit motives, as well proposed alternatives to vaccines. 
Topic 2 has ``big'', ``pharma'', ``trust'' and ``rushed' as highly scoring words, with many users questioning whether the vaccine is pushed primarily to make money: ``Big Pharma can’t make as much money with a pill versus a year long process for a vaccine.''. Many users voice their concerns about the vaccines directly to the then-sitting president, tagging ``@realDonaldTrump'' in their tweets. Another commonly raised critique in the discourse concerns the fact that vaccine manufacturers can not be easily sued, seen in topic 28: ``Who would want a vaccine when Drug companies can't be sued if something goes wrong there has not been enough testing of vaccines to make sure they are safe.''. Two topics (15 \& 33) mostly discuss hydroxychloroquine as a better measure than vaccines against COVID-19. Topic 25 is a topic where people discuss the protection antibodies give against the virus. A commonly raised argument is that the vaccines should not be necessary for those who already had the virus. Topic 31, a topic with 624 tweets, is dominated by tweets citing a high recovery rate from COVID-19 as an argument against the vaccines. An example of a very typical tweet from this category is: ``If there is a 99\% recovery rate, why would we need a vaccine?''. The topic grew in popularity during the second half of 2020, reaching a peak in the months leading up to the vaccine roll-outs, see Figure \ref{figure:pharma_hcq_99_hcq}.

\subsection*{Popular Conspiracies (topics 12 and 13)}
\label{5G}
\begin{figure*}[h]
  \centering
  \includegraphics[width=\linewidth]{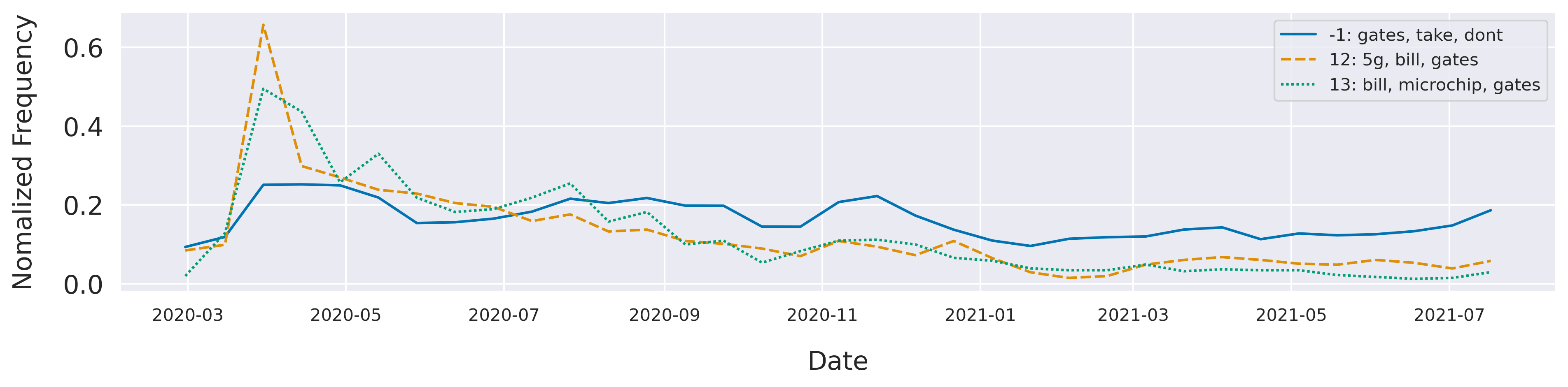}
  \caption{Bill Gates, 5G, and microchip topics over time.}
  \label{figure:5g}
\end{figure*}

Topic 12 has a very strong peak in April (see Figure \ref{figure:5g}) which coincides with the first 5G network tower to be put on fire on the second of April in Liverpool, UK \cite{satariano_burning_2020}. 
Apparently a rumor started that 5G towers are partially to be blamed for the spread of COVID-19. These rumors were combined with the sentiment that the 5G towers are used to control people's minds using microchips that are supposedly inserted through the Corona vaccination.
The Major of Liverpool declared these rumors as false, which seems to have triggered some people to take action into their own hands. 
This incident triggered discussions among the vaccine-hesitant population, as can be seen in our timeline, where the words ``5G'', ``tower'', ``chips'', ``microchips'' are highly weighted. 

Topic 13 is closely related to the topic of 5G (topic 12) but with a stronger focus on the microchip rumor. The peak and curve behaviors are very similar for both topics (as they both share microchip as an important word). They peak in the beginning of April and after that lose in importance. The peak is co-occurring with the following events that may be contributing to its peak. On March 18, Bill Gates logged onto Reddit and answered questions. While doing so, he predicted, that one day we would all carry a digital passport of our health records. He did not suggest a microchip for that but some kind of e-vaccine card. On March 19, a Swedish site picked that up and put the following headline to their article. ``Bill Gates will use microchip implants to fight coronavirus.''. With that the conspiracy theory was born. The discussion it started among vaccine-hesitant people is reflected in our data. Simultaneously the non-governmental institution ID2020 (Digital Identity Alliance) was brought into this as they are advocating for a digital ID for undocumented people like refugees. Vaccine hesitant individuals drew the conclusion that ID2020 is involved in putting microchips into people to reach their goal. Topic 13 has words like ``gates'', ``implant'', and ``tacking'' highly rated with many talking about how the pandemic was planned by gates: ``It's already planned by Gates. Thank Trump for not allowing ANY Gates vaccine!''. The topic did not closely correlate with Google search volume for Gates, indicating that this could be a topic with specific popularity on Twitter.

\subsection*{Testing and Guinea pigs (topic 8 and 16)}
This topic mostly contains tweets talking about the vaccine trials with a critical tone. The terms ``pig'' and ``pigs'' score highly in this topic because of their appearance in the phrase ``Guinea pigs'', used to refer to those choosing to take the vaccine. One user said: ``I WILL NOT BE A GUINEA PIG with this vaccine! Why don’t you and your family be the test dummies !''. The topic had most popularity during the vaccine trials of 2020, and has since lost traction during the roll-out of the vaccines. 

Closely related is the topic of testing (topic 16) that also follows a similar trend over time. Important keywords for this topic like ``rushed'', ``tested'' and ``wouldn't'' reflect the stance expressed that the development process is being ``rushed'' and an unwillingness to take an ``untested'' vaccine. ``I have to agree with you on not taking it. This is an example of why vaccines should not be rushed with much needed testing not done before it's approval.''.

\subsection*{Pfizer and AstraZeneca (topic 10)}

\begin{figure*}[h]
  \centering
  \includegraphics[width=\linewidth]{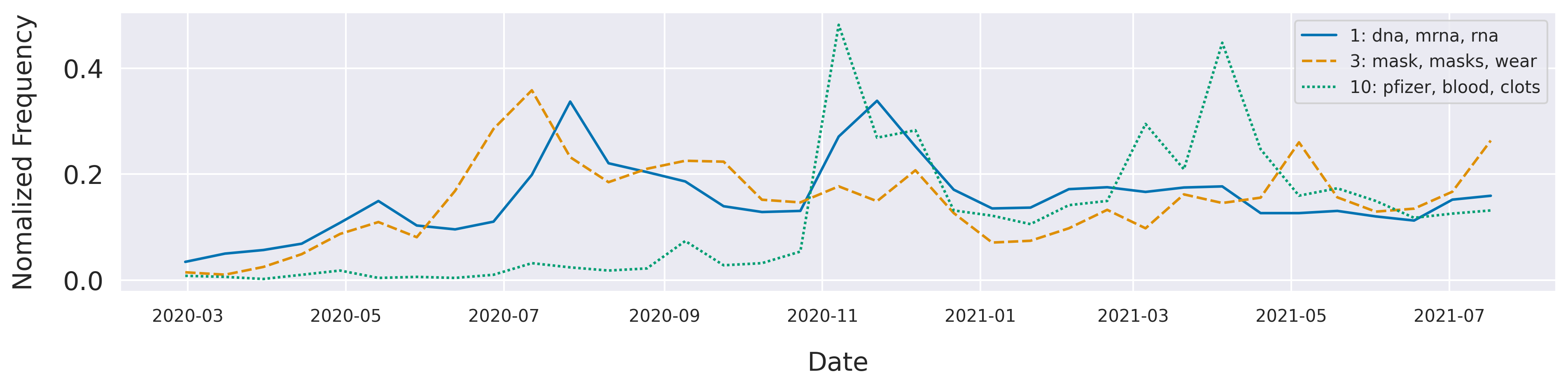}
  \caption{Pfizer, mRNA, and mask topic over time.}
  \label{figure:pfizer_mrna_masks}
\end{figure*}

This topic peaks in November 2020, see Figure \ref{figure:pfizer_mrna_masks}. Compared to Google Trends, there is a clear peak with the search term for Pfizer and Corona on November 9. This is the day that Pfizer announced a successful trial 3 phase of their corona vaccine with an effectiveness of 90\%. On the same day ``Nature'' and ``BBC News'' reported on it. This announcement likely lead to an increase of critical voices concerned about the safety of the vaccine and triggered discussions among vaccine hesitant groups. 
We also see peaks of discourse during the month of March 2021, with the key term ``aztrazeneca'' and ``blood clots'' being used especially often during this period. At this time concerns surrounding a potential side effect of blood clots started circulating, causing many countries to pause their use of the AstraZeneca vaccine. These news seem to have caused a lot of negative discourse around the vaccine, such as this user voicing their distrust in the process: ``Denmark, Norway, Iceland and Bulgaria halt use of AstraZeneca’s COVID-19 vaccine over reports of blood clots. Has any research been done on effects on various blood types and regional DNA variances in countries?? Thought NOT. Don't trust Chinese ``run'' WHO!''.

\subsection*{DNA and mRNA (topic 1)}
This topic has two peaks in the investigated time-line. The first peak can be seen in August 2020 (see Figure \ref{figure:pfizer_mrna_masks}). Pfizer published its results for phase 1 and phase 2 clinical trials for its mRNA-based vaccine in the journal Nature on August 12 \cite{mulligan_phase_2020}. Around this time, many news outlets started to describe the promising vaccine candidates, with a special focus on the new technique to use mRNA, as done by Pfizer and Moderna. This new technique to use the body's own protein generating mechanism to produce the spike protein of the coronavirus instead of using a dead or a weakened virus, gave rise to concerns about the vaccines' safety. Some people were concerned that it could have additional, adverse effects to a human's DNA and permanently alter it. The second peak of the mRNA topic in December 2020 lies close to a peak also seen for the Pfizer topic, likely due to the publication of Pfizer's phase 3 trails utilizing the mRNA-technique.

\subsection*{Preventative Measures (topic 3, 6, and 32)}
One of the largest topics found in the dataset with 8,984 tweets concerns face masks as a protection measure against the virus. The topic picked up pace in June 2020 (see Figure \ref{figure:pfizer_mrna_masks}), and remained popular through the entire period of study. Top-words for this topic include ``mask'', ``wearing'', ``distancing'', ``protect'' and ``mandatory''. Since the analysis is limited to tweets with negative stance toward vaccines, much of the discourse consists of critiques against mask mandates, calling them discriminatory. One user said: ``Yet if I don't take the covid ``vaccine'' or wear a mask you want me to suffer total discrimination.''. Other users expressed support of masks seeing it as an alternative to vaccines: ``Definitely masks but no to the vaccine. I don't trust it because djt pushed them too fast.''.

In addition to masks, lockdowns is another measure that has been widely used in an attempt to limit the spread of the virus. Many users in topic 3 are questioning policies being set in place to limit the spread of the virus. Tweets from early in the pandemic are often arguing that lockdowns are unfeasible with no end in sight for the pandemic. As the vaccines were deployed, this topic seems to have shifted arguments that the lockdowns have been put in place in order to persuade people to take what they perceive as an unnecessary vaccine. 

The eighth-largest topic of the model concerns the vaccination of children (topic 6). A commonly expressed view in this topic is that the vaccine is too experimental to use on children and that it is unethical to make vaccines mandatory for attending school. Many parents claim that they will opt for homeschooling rather than a vaccine. An argument often used in this topic is that kids have strong immune systems, and should therefore not require a vaccine. Another common sentiment in this topic is that the fear of COVID-19 is overblown and that schools should be re-opened.

\subsection*{Parallels to other disease (topics 17, 19 and 20)}
Three topics make comparisons between viral infections and COVID-19. The three most common in our model are Polio, HIV and earlier strains of SARS. Many raise the argument that it is implausible that an effective vaccine has already been found for COVID-19, since many years of research never lead to a successful vaccine for HIV or SARS. Many also question mRNA technology, arguing that it is not a ``real vaccine''. Some users bring up the Polio vaccines as an example of vaccines that actually worked, since this disease has been eradicated in most places, something that cannot be said for COVID-19.



\section{Discussion}
\subsection{Principal Findings}
In this research we investigated the discourse about COVID-19 vaccines that took place in English on Twitter between March 1, 2020 to July 31, 2021. We developed a classifier to identify tweets with a negative stance toward vaccines, and then modeled the topics of the negative tweets. Looking at the topics found, we see that a number of different conspiracy theories play a large role in the negative discourse. Some topics have an obvious link to popular conspiracy theories, such as those about 5G towers, microchips, and Bill Gates. However, in topics discussing other concerns such as COVID-passports, pharmaceutical companies, and racism, references to grand conspiracies are also commonplace. Another overarching theme concern negative perceptions of how restrictions and guidelines impact daily life. Distrust in pharmaceutical companies also seems to fuel much of the hesitancy, and a discussion of alternatives to vaccines. 
 
 The percentage of negative tweets, as well as these more conspiratorial topics, saw a noteworthy increase in the month following the declaration of the pandemic. It appears that the attention put on the pandemic gave fuel to these conspiracies, that previously held a more fringe position. Previous research has indicated that anxious people are more likely to believe in and share misinformation \cite{freiling_believing_2021, anthony_anxiety_1973}. It could therefore be hypothesized that an increase in anxiety provoked by the news of a pandemic could explain the increase of hesitant and conspiratorial tweets in the following months. Furthermore, partaking in misinformation sharing has also been shown to contribute further to anxiety \cite{verma_examining_2022}. It could therefore be posed that a vicious circle could be formed where anxiousness leads people to share more misinformation, leading to even more anxiety. Another possible explanation is that these negative ideas were able to spread to a higher degree as the network of people involved in the discussion grew.
 
 Similar to \citet{lyu_covid-19_2021} we saw a general decline in the percentage of negative tweets co-occurring with the vaccine roll-out. Furthermore, beyond the time-frame investigated by \citet{lyu_covid-19_2021}, we could show that negativity stayed on this lower level during the following seven months. 
 
A decreased popularity was also observed for many of the more conspiratorial topics such as 5G towers, microchips, and the Gates foundation. This could be indicative that communication by governments and health authorities around the issue was successful at combating some of the negative perceptions people had about the vaccines. However, another explanation could be that the deployment of vaccines attracted more public attention, shifting the composition of users involved in the discourse. It can also not be ruled out that Twitter started enforcing their crisis misinformation policy more strictly as the vaccination process began, possibly contributing to a decrease in conspiratorial and vaccine-hesitant tweets. 

\subsection{Limitations}
It is important to note that this study is limited in its scope to only English-speaking Twitter. Although \citet{Sloan2015} have found that there is widespread Twitter use among the general population, the conversations observed in this study cannot be considered representative of the entirety of the general public discourse. Furthermore, the demographic using Twitter in English likely also differs in many respects to the overall user-base. The analysis of topics was also limited to negative discourse. Future studies could benefit from investigating how the positive discourse has changed over the course of the pandemic. This could also help answer questions about which positive discussions took place as the vaccines rolled out, taking up space previously held by negativity. As the pandemic has continued into 2022, new studies could also include an even longer time period to get a more comprehensive picture of the evolution of topics over the course of the pandemic 

\subsection{Conclusion}
Although the negativity toward vaccinations on Twitter decreased as the vaccines rolled out, many countries still faced difficulties with vaccination willingness within their populations. In our analysis of negative tweets over the course of the pandemic, we saw different indications for vaccine hesitation such as low trust toward authorities, strong insistence on personal freedom with respect to following guidelines such as the wearing of masks or uptake of vaccines, as well as a non-negligible influence of conspiracy theories around COVID-19 itself as well as the vaccinations. While the percentage of negative tweets and conspiratorial topics reduced as the vaccines rolled out, it is still possible that the negative discourse in the preceding months had already swayed some users into a more hesitant stance. As unvaccinated people face a much higher rate of hospitalization and death \cite{dyer_covid-19_2021}, hesitancy could have severely negative outcomes on society. In a country where non-pharmaceutical interventions are relaxed, it has been estimated that hesitancy could lead to a more than 7 times higher mortality \cite{mesa_modelling_2022}. 
In order to improve the handling of the current pandemic as well as being prepared for future pandemics, a successful communication strategy should address concerns circulating on social media in an early stage, preventing negative perceptions from taking hold. The fact that such bizarre conspiracies as mind-controlling 5G antennas were widely circulated must be seen as a failure of communicative and educational efforts and take up valuable space where constructive conversations around the pros and cons of vaccination could be had.

\vspace{6pt}

\subsection*{Data Availability}
The dataset of tweets annotated for the purpose of this study has been made available on GitHub \cite{dataset}. Plots can be viewed interactively on the companion website \cite{companion_website}.
\subsection*{Acknowledgments}
We acknowledge the computational resources provided by the Aalto Science-IT project. We also acknowledge the input from Leila Gharavi, the feedback from Dr. Koustuv Saha, as well as the work of the two annotators. 
\subsection*{Author contributions}
Conceptualization, G.L., T.A. and B.K.; methodology, G.L., T.A. and B.K; software G.L.; validation, G.L., T.A. and B.K; formal analysis, G.L., T.A. and B.K.; data curation, G.L.; writing---review and editing, G.L., T.A. and B.K. All authors have read and agreed to the final version of the manuscript.
\subsection*{Funding}
This research received no external funding.
\subsection*{Conflicts of Interest}
The authors declare no conflict of interest.



\newpage
\bibliographystyle{abbrvnat}
\bibliography{ms}  






\end{document}